# Strain-gradient induced topological transition in bent nanoribbons of the Dirac semimetal $Cd_3As_2$

Wen-Zhuang Zheng, Tong-Yang Zhao, An-Qi Wang*, Dai-Yao Xu, Peng-Zhan Xiang, Xing-Guo Ye, and Zhi-Min Liao*

*State Key Laboratory for Mesoscopic Physics and Frontiers Science Center for Nano-optoelectronics, School of Physics, Peking University, Beijing 100871, China*

\* anqi0112@pku.edu.cn (A.-Q. Wang); liaozm@pku.edu.cn (Z.-M. Liao)

**Abstract:**

Dirac semimetal is an ideal parent state to realize various exotic states of matters, such as quantum spin Hall state, Weyl semimetal phase and Majorana zero modes. Topological phase transition allows for the switching between these different topological states. Here, in this paper, we exhibit experimentally an effective approach of inducing topological phase transition in $Cd_3As_2$ nanoribbons, by applying a bending strain profile onto the sample. The local strain varies linearly from compression to tension through the cross-section of a bent nanoribbon. The strain gradient causes obvious lattice deformation and breaks the $C_4$ rotational symmetry, thus opening an energy gap at the Dirac points and making the bulk gapful. When further increasing the strain strength, the local strain effect dominates over the symmetry-breaking effect, where spatially-varying band shift becomes prominent across the nanoribbon. Our results demonstrate the effect of strain gradient on the evolution of energy band structures, which should be valuable for further study of strain-mediated topological phase transition.

## INTRODUCTION

The combination of topology and physics has ushered a variety of quantum materials, such as topological insulators [1,2], Dirac semimetals [3-6], Weyl semimetals [3-6] and topological superconductors [1]. Topological Dirac semimetals (TDSs), featured with gapless Dirac cones in the bulk band structure, are of vital importance as a parent state of various other topological phases. For instance, reducing the material thickness would possibly trigger a phase transition from TDSs to topological insulators [7,8]. The quantum confinement effect in an ultrathin TDS film would lead to a crossover to a two-dimensional topological insulator, with the emergence of quantum spin Hall state [8,9]. Upon breaking the time-reversal symmetry via external magnetic field or magnetic exchange interaction, the original Dirac cone would split into Weyl cones, accompanied by the presence of magnetic Weyl semimetal phase [3-6]. Besides, when proximitized with s-wave superconductors, TDS is proposed to support topological superconducting phase and Majorana zero modes [10-13], which are significant for implementation of topological qubit and fault-tolerant quantum computation. TDS provides an ideal platform to study the phase transition between different quantum states, which is valuable for fundamental and application-based studies and has attracted current research interest.

As a prototypical TDS, $Cd_3As_2$ has two gapless Dirac cones along the $k_z$ direction protected by the $C_4$ rotational symmetry [8]. Pristine $Cd_3As_2$ has demonstrated abundant novel features, including chiral anomaly [14], π Aharonov-Bohm oscillations [15,16], quantum Hall effect [9,17,18] and spin-polarized Fermi arc surface states [19-21]. In previous literatures on topological phase transitions in $Cd_3As_2$, widely-acknowledged useful methods include introducing interplay with external electromagnetic fields [22-24], or reducing sample dimensions to introduce quantum confinement effect [8,9]. Experimental works have demonstrated feasibility of magnetic field-driven topological phase transition in $Cd_3As_2$ into Weyl semimetals, quantum Hall insulators, or trivial insulators [9,14,17,18,25,26], and predictions that Dirac semimetal $Cd_3As_2$ can turn into a 2D topological insulator in an ultrathin film are also proposed [8]. Recently, an emergent type of mechanism that induces topological phase transitions in topological semimetals has been proposed, that is, introducing certain strain profiles onto the whole lattice [27-30]. Rather than introducing external electromagnetic fields or reducing the sample thickness, the strain-induced band modulation can be easily achieved during sample fabrication, which is more static and intrinsic as a sample property, and more intuitive in the symmetry analysis. The correlation between strain profile and gauge field theories further enriches the physical implication of relevant topological phase transitions [31-34].

In this work, we demonstrate the variation of electronic properties modulated by static bending strain applied on $Cd_3As_2$ nanoribbon devices through transport experiments. Compared to the case of straight nanoribbons, the bent nanoribbon displays some distinct features, *i.e.*, a second turning point in the $\rho-T$ curve and flat region in the $\rho-V_g$ curve. This can be well understood by considering that the introduced strain gradient breaks the pristine lattice symmetry and opens a global bulk gap near the neutrality point. When further increasing the bending strength applied onto the nanoribbon, the local strain induced energy band shift dominates and renders the bulk gap closing.

**METHODS**

$Cd_3As_2$ nanoribbons were grown by chemical vapor deposition method as reported in our previous work [35]. $Cd_3As_2$ nanostructures with a typical scale of several tens of microns in length and several hundreds of nanometers in diameter were obtained and examined using scanning electron microscopy (SEM), as shown in Fig. 1(a). The nanoribbons show excellent mechanical flexibility, allowing us to investigate the effect of bending profile on them without introducing fractures to the samples. High-resolution transmission electron microscopy (TEM) image shows the nanoribbon has an interplanar spacing of about 0.44 nm, indicating the [110] growth orientation [Fig. 1(b)]. In this work, the selected nanoribbons have a thickness ranging from 170 to 200 nm. During device fabrication, the nanoribbon was first transferred onto the Si/SiO$_2$ substrate, and an in-plane bending profile was then introduced mechanically under an optical microscope using a glass tip. To help consolidate the bending shape of nanoribbon, a 3-nm-thick $Al_2O_3$ layer was deposited on the whole sample using atomic layer deposition (ALD) system. Standard electron beam lithography and electron beam evaporation techniques were adopted to fabricate Ti/Au electrodes on regions of nanoribbons with variant bending curvatures. We have experimentally investigated a total of 12 bent nanoribbon devices, in which the results are consistent with each other. In the following, we mainly elaborate on the three typical devices, marked as device A, B and C (see Appendix A).

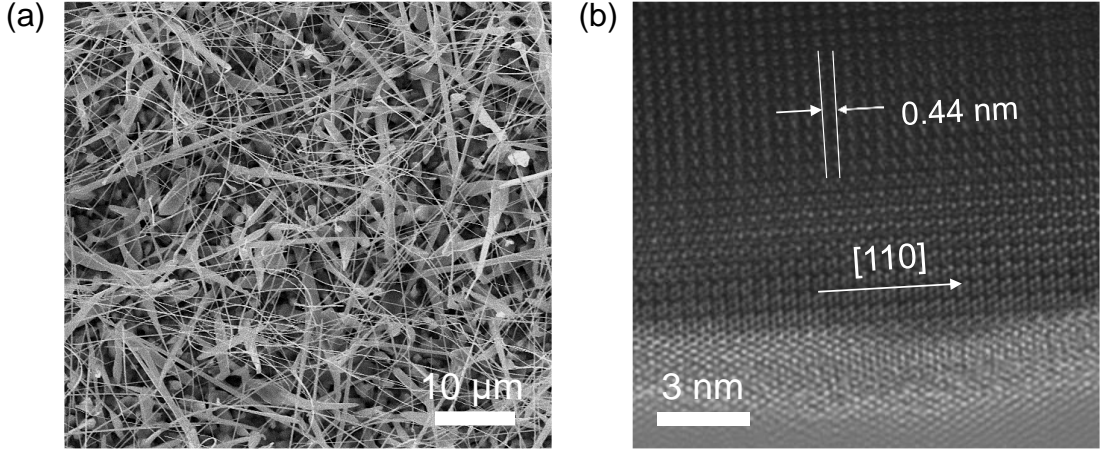

**FIG. 1. Characterization of the synthesized Cd₃As₂ nanoribbons.** (a) SEM image of the as-grown $Cd_3As_2$ nanoribbons, demonstrating great flexibility. (b) High-resolution TEM image of a typical nanoribbon with interplanar spacing about 0.44 nm, indicating the [110] growth orientation.

Figure 2(a) shows the SEM image of a bent $Cd_3As_2$ nanoribbon device. In such bent nanoribbon, lattice deformation and strain mainly happen along the axial direction of the ribbon. The strain $u$ varies linearly from tension (outside the ribbon) to compression (inside the ribbon) across a strain-neutral plane (geometrical middle-plane of the ribbon), leading to a strain-gradient along the radial direction [Fig. 2(b)]. The value of strain-gradient is $g = 1/r$, where $r$ is the local curvature radius. Then, the maximum values of tensile and compressive strain at the outer and inner side can be estimated by $\varepsilon = \pm\frac{d}{2r}$, where $d$ is the width of nanoribbon, as illustrated in the top view in Fig. 2(b).

From the above perspective, the lattice deformation is varying spatially in a continuous manner, making the system a complex "parallel" configuration of many local lattice structures. Thus, a theoretical analysis of band evolution is necessary here to shed light on the transition of transport properties in a bent nanoribbon, serving as a guidance for experimental exploration. The topological semimetal nature of unbent $Cd_3As_2$ arises from the inverted bands due to strong spin-orbit coupling [8], as depicted in Fig. 2(c). The strength of coupling, as well as all the hopping amplitudes in the system, should be functions of lattice parameters. Therefore, the strain profiles applied on $Cd_3As_2$ nanoribbons in our experiment generate corrections to local lattice parameters, and directly changes the low energy band structures. By unidirectionally increasing or decreasing certain lattice parameters, the conduction and valence band shall accordingly move, as indicated by relevant DFT calculations [36,37], towards a lower or higher energy. In the case of a weak bending profile, a considerable energy gap induced by the breaking of $C_4$ rotational symmetry shall emerge [8,25,37,38], while the overall energy shifts of bands remain sufficiently small. There may exist a global energy gap in the whole sample, and the system would exhibit insulating behavior when the Fermi level is tuned into the global gap [Fig. 2(d)]. In the strongly bending case, however, the shift of energy band becomes more significant, and no universal gap for the whole sample exists [Fig. 2(e)]. Thus, the Fermi level always intersects with energy bands, and the corresponding transport measurements manifest a p-type to n-type crossover in one single device. We believe that the bending gradient $g$, which determines the size of gap induced by symmetry reduction, and the sample width $d$, which relates to the maximal band shift, together decide the final energy band structure of the system.

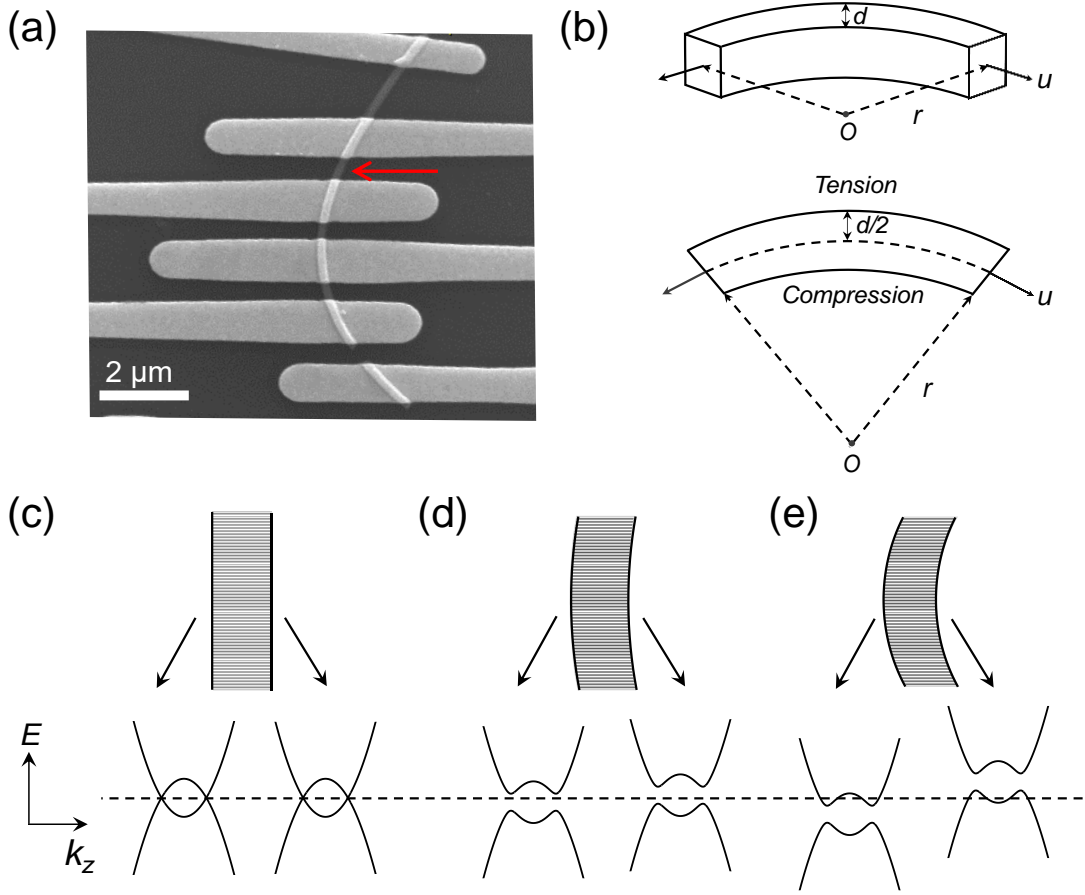

**FIG. 2. Schematic of bending strain profiles and corresponding band structures.** (a) SEM image of the bent nanoribbon device A. The red arrow indicates the position A1 (local curvature radius $r = 4.13$ μm, strain-gradient $g = 24.2\%\ \mu m^{-1}$) for transport measurements in Fig. 3. (b) Schematic illustration of the bending strain acting on the Cd$_3$As$_2$ nanoribbon. The strain $u$ along the axial [110] direction changes from tension (outside the ribbon) to compression (inside the ribbon) across a strain-neutral plane (geometrical middle-plane of ribbon). The nanoribbon width and radius of curvature are marked as $d$ and $r$, respectively. (c-e) Schematic of bulk band structure evolution in the bent Cd$_3$As$_2$ nanoribbon as the bending strength increases. Quantum confinement is ignored here for simplicity. (c) Case of the straight nanoribbon. Cd$_3$As$_2$ has two gapless Dirac cones along the $k_z$ direction protected by the C$_4$ rotational symmetry. Absence of strain insures the band structure on both sides of sample to be identical. (d) Case of weak bending strain. On the one hand, the strain gradient reduces lattice symmetry and opens a bulk gap, resulting in the "$\omega$" shape of local band

structure. On the other hand, local strain modifies lattice parameters, which further induces slight energy shift in conduction and valence bands. (e) Case of strong bending strain. With the band shifts further, the system will enter an "indirect gapless" regime, and the Fermi level will always intersect with conduction or valence bands.

**RESULTS AND DISCUSSION**

The results discussed in the main text all come from the device A. Transport measurements were performed in three different bent regions of device A, *i.e.*, region A1, A2 and A3 (see Appendix B). We firstly investigate on the bent region A1 (denoted by the red arrow in Fig. 2(a)). The local curvature radius $r$ and width $d$ of the bent region are 4.13 µm and 199 nm, respectively, corresponding to a maximal strain amplitude $\varepsilon = 2.41\%$. The $\rho - T$ curve corresponding to A1 is plotted in Fig. 3(a). The $\rho - T$ curve exhibits a clear insulating behavior in the high temperature region, where resistivity $\rho$ increases monotonically with decreasing $T$. Such kind of insulating behavior of $\rho - T$ curves in high temperature region has actually been observed in several semimetallic systems with high sample quality, including $Cd_3As_2$ (Ref. [39]) and $ZrTe_5$ (Ref. [40]) for instance, which is not the point of novelty here in our results. The more noteworthy feature is the second turning point which emerges at a low temperature of around 10 K (see inset of Fig. 3(a)). In previous reports on $\rho - T$ curves of topological semimetals, the resistivity increases to only one single maximum with decreasing temperature, then drops monotonically until the temperature approaches 0 K. In our result, the emergent second turning point at 10 K is a very distinctive phenomenon, and has implication on possible phase transitions occurring in the bent $Cd_3As_2$ nanoribbon. Especially, the insulating $\rho - T$ behavior below 10 K may indicate the opening of a bulk gap in the energy band structure. The metal-insulator transition has been confirmed from $\rho - T$ spectrum in many systems [41-47]. We also performed measurements on gate voltage dependence of the resistivity in the bent region, as shown in Fig. 3(b). Compared to the standard $\rho - V_g$ curves obtained in the straight $Cd_3As_2$ nanowires [48], a conspicuously different feature can be observed here, that resistivity maximum no longer exhibits a single peak lineshape, but rather a plateau-like flat region, with plateau width approximately 20 V. The stable maximal resistivity value over a relatively wide range of gate voltage is consistent with the scenario of an energy gap in the bulk band structure [49-51]. Once the Fermi level is tuned into the gap, the bulk conduction is greatly suppressed and the overall conduction becomes dominated by other mechanisms, including disorders and surface states [52-54], which are little affected by gate voltage. Therefore, a relatively flat region between n-type and p-type conduction in $\rho - V_g$ curve is experimentally observed here. In contrast, for an ideal Dirac semimetal system, the gapless band structure results in a sharp transition between n- and p-type conduction, and the resistivity near the Dirac point should therefore take the form of a narrow peak. Electron and hole mobility can also be extracted from $\rho - V_g$ curve, which take the values $\mu_e = 1.59 \times 10^4$ cm$^2$/(V · s) and $\mu_h = 3.81 \times 10^2$ cm$^2$/(V · s), obviously reduced from the typical values in high-quality Cd$_3$As$_2$ nanostructures [55]. The flat region in our $\rho - V_g$ curve as well as the reduced mobility thus reveal from a parallel aspect the breakdown of Dirac cone structure in the bent nanoribbon.

To better reveal the topological transition occurring in the bent nanoribbon, we also perform magnetotransport measurements on the bent region under an out-of-plane magnetic field $B$. Figure 3(c) shows the resistivity variation $\Delta\rho$ as a function of $1/B$ in the bent region A1. The oscillatory pattern here is reminiscent of Shubnikov-de-Haas (SdH) oscillation, which emerges from quantized Landau levels under high magnetic field. In our result, the oscillation seemingly exhibits a periodic behavior under high magnetic field $B > 5$ T ($1/B < 0.2$ T$^{-1}$), allowing us to derive the Landau level index plot $N - 1/B$, as shown in Fig. 3(d). Here, oscillation peaks correspond to integer indices, and valleys to half-integer indices. An obvious linear relation in $N - 1/B$ is clearly observable, and the intercept on $N$ axis is approximately $-0.6$. Similar SdH oscillation and Landau fan diagram are also captured on the small strained region of other devices, *e.g.*, device B (see Appendix C). It is worth noting that, according to the gate voltage dependence curve [Fig. 3(b)], the Fermi level of the system is situated within the bulk gap at zero gate. Therefore, the observed SdH oscillation pattern without gate modulation should not come from the formation of bulk Landau bands. A highly possible origin of the observed oscillation might be the topological surface states. Further, the obtained intercept $-0.6$ is close to the theoretical value $\pm 1/2$ expected from topological insulator surface states [56-61], implying that the tendency of transition into topological insulator phase is happening in the bent nanoribbon sample. Actually, previous theoretical calculations [7,8,62] have predicted that due to the inverted band structure, Dirac semimetal $Cd_3As_2$ would be driven into a 3D topological insulator under symmetry breakings. In our experiments, the applied bending strain serves as the symmetry-breaking factor and would break the $C_4$ rotational symmetry, rendering the presence of topological insulator phase.

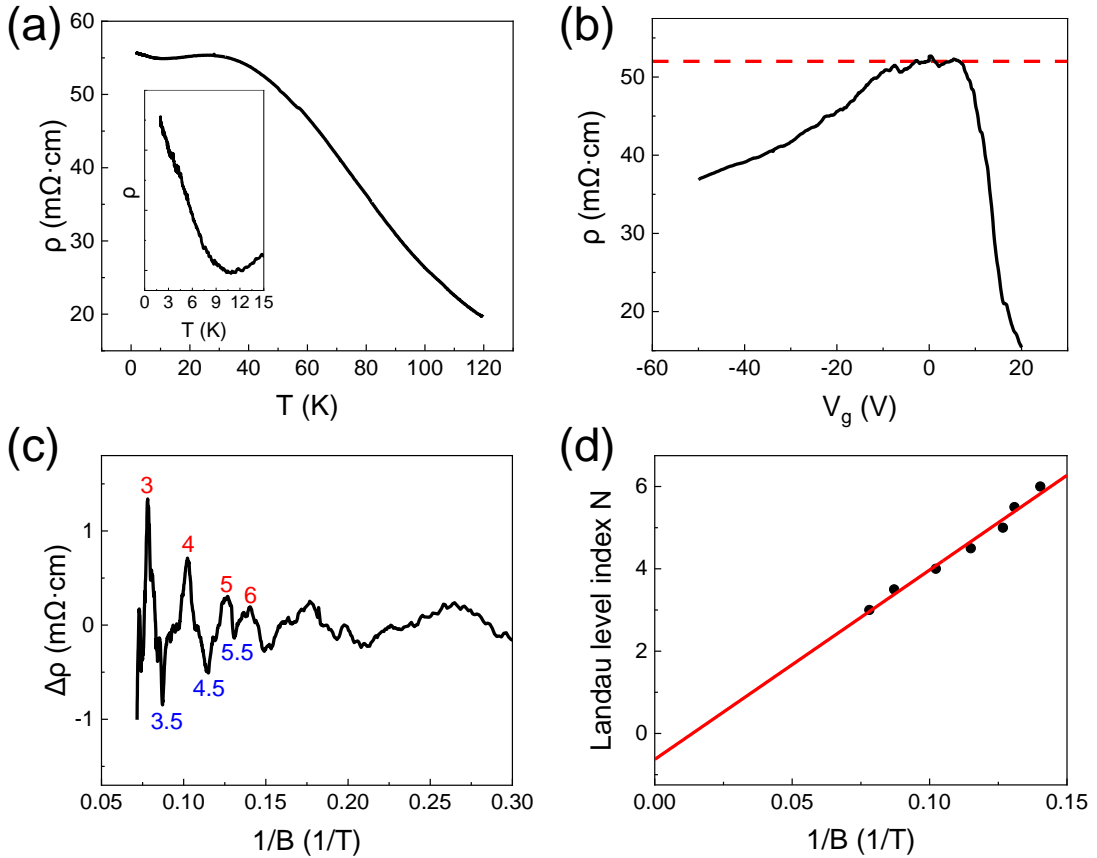

**FIG. 3. Transport measurements on the bent region A1.** (a) Measured resistivity $\rho$ as a function of temperature $T$. Inset: zoom-in view of $\rho - T$ plot around 10 K. (b) Gate voltage dependence of resistivity. A flat region is observed between $V_g = -10$ V

and $+10$ V with a dashed guide-to-the-eye curve. (c) SdH oscillations are observed. Landau levels are labeled with oscillation peaks being integers and oscillation valleys being half integers. (d) Landau level index $N$ plotted against $1/B$. Peak and valley values obtained from SdH plot (c) are specified with black dots.

According to our general analysis of band evolution in the bent $Cd_3As_2$ nanoribbon system with large strain, local lattice deformation varies notably with respect to spatial position due to the inhomogeneous strain profile, making the band energy shift the dominant effect in this case. In previous literatures, there were similar reports concerning the drastic changes in band profiles caused by large strain in topological materials, for example, in HgTe system [63,64]. In the complex system of bent $Cd_3As_2$ nanoribbon here, it is reasonable to anticipate consecutive new transition of transport properties by further increasing the strain and exceeding the weak symmetry-breaking regime. To investigate on the corresponding possibility, we performed transport measurements on a second region, noted as A2 [Fig. 4(a)], of the same nanoribbon sample, where strain was designed to be much enhanced than region A1. In the bent region A2, the maximal strain is estimated to be $\varepsilon = 2.80\%$ with $r = 3.46$ μm and $d = 194$ nm, respectively.

Gate-voltage dependence of the resistivity in the largely-bent nanoribbon region A2 was measured, as shown in Fig. 4(b). Interestingly, it is observed that the typical transfer curve of a gapless semimetallic system, featured with a single sharp resistivity peak, again emerges in the largely bent nanoribbon. From the transfer curve, we can obtain the carrier mobilities $\mu_e = 2.68 \times 10^4$ cm$^2$/(V·s) and $\mu_h = 7.32 \times 10^2$ cm$^2$/(V·s), much enhanced compared to the weakly bent case A1. Resistivity oscillations are also observed for the largely-bent region, as shown in Fig. 4(c). The collective oscillation is a combination of multiple SdH oscillation patterns with different frequencies, as indicated by the fast Fourier transformation (FFT) spectrum in Fig. 4(d). The multiple-frequency oscillation pattern is also observed in the bent region A3 of the same nanoribbon, which has the similar bending strength and strain gradient as those of region A2 (see Appendix B).

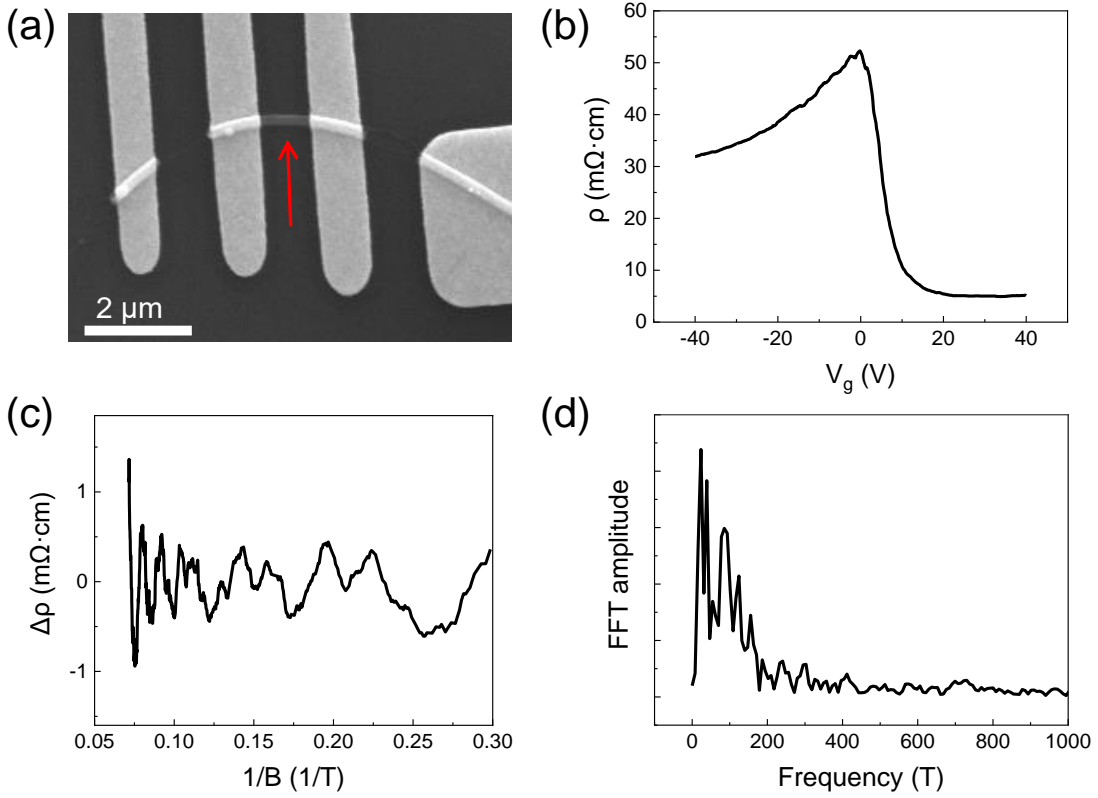

**FIG. 4. Transport measurements on the largely-bent region A2.** (a) SEM image of the region A2 from the same nanoribbon as that in Fig. 2(a). The red arrow indicates the measurement region A2. The local curvature radius is $r = 3.46$ μm, corresponding to a maximal strain amplitude $\varepsilon = 2.80\%$. (b) Gate voltage dependence of resistivity. Single sharp peak feature typical for a semimetal is retrieved. (c) The resistivity oscillation $\Delta\rho$ after subtracting the background. (d) FFT spectrum of resistivity oscillation in (c).

Within the framework of band evolution in the bent $Cd_3As_2$ nanoribbon as we proposed earlier [Figs. 2(c-e)], the experimental results observed here can be well understood. First, the gapless Dirac semimetal feature seems to be retrieved in the gate-voltage modulation result [Fig. 4(b)]. This can be understood as follows. For simplicity, we regard the conduction in nanoribbon as a parallel connection of two sample edges, as depicted in Fig. 2(e). When Fermi level is raised by gate, it will first touch the valence band top and then enter the energy gap on the outer side of ribbon (lower-left panel), while the inner side of ribbon still has the Fermi level crossing its valence band (lower-right panel). In this case, the conduction is dominated by the inner side of ribbon, corresponding to p-type conduction. By further raising the Fermi energy, it will reach conduction band bottom of the outer side in priority, then reach the valence band top on the inner side. In such a procedure, conversion from p-type to n-type happens, which can be a very rapid change in analogous to typical metallic system. By stepping further on raising the Fermi level, the inner side become insulating while the outer side showing dominant n-type conductance. In the real nanoribbon system where infinite folds of band configuration are parallelly connected, the main idea in our simplified model should still hold true and applicable to explain the obtained results. Then the resistivity oscillation in Fig. 4(c) can be explained in the similar manner as well. The conduction in bent nanoribbon comprises of contribution from multiple different lattice configurations with varying lattice constants and different band profiles. Since Fermi level is equal everywhere in the sample, one can expect that cross-sectional area of "local Fermi surface" is also spatially dependent. Therefore, the observed resistivity oscillation in A2 and A3 is actually a superposition of many SdH oscillations with different frequencies, according to Lifshitz-Onsager quantization rule [65]. The observation of a multi-frequency SdH oscillation pattern thus also verifies the prominent band evolution induced by strong local strain effects, in accordance with the general comprehension proposed by us. The strain-mediated SdH oscillations are also observed in device C, where single-frequency SdH oscillation emerges in the weak strain regime and multi-frequency oscillation in the strong strain case (see Appendix D and E). Notably, the Fermi level of nanoribbon in device C is far away from the charge neutrality point, different from the case of device A and B.

**CONCLUSION**

In conclusion, we have performed detailed investigation on the effect of bending strain profile on the transport properties of bent Dirac semimetal $Cd_3As_2$ nanoribbon. The introduced strain gradient would break the pristine lattice symmetry and result in the gap opening near the neutrality point. When further enhancing the strain strength, the gap gradually closes, which possibly originates from nonnegligible local strain effect that dominates over the symmetry breaking effect. Our results demonstrate the strain-gradient induced topological phase transition in Dirac semimetals via transport methods. This approach to induce phase transition should be generic not only to Dirac semimetals, but also to other topological materials.

**ACKNOWLEDGEMENTS**

This work was supported by the National Natural Science Foundation of China (Grant Nos. 91964201, 61825401, and 11774004).

**APPENDIX A: CHARACTERIZATION OF THE FINISHED BENT NANORIBBON DEVICES**

Figure 5 shows the optical image of three typical devices, marked as device A, B and C, respectively. The red arrows indicate the bent regions with different strain strength of a same nanoribbon sample. The thickness of selected nanoribbons ranges from 170 nm to 200 nm, which ensures Dirac semimetal phase in unbent conditions. The local curvature radius $r$, strain-gradient $g$ and maximum strain amplitude $\varepsilon$ of these bent regions are summarized in Table 1. The value of strain-gradient is $g = 1/r$, and the maximal strain amplitude can be estimated as $\varepsilon = d/2r$, where $d$ is the width of nanoribbon.

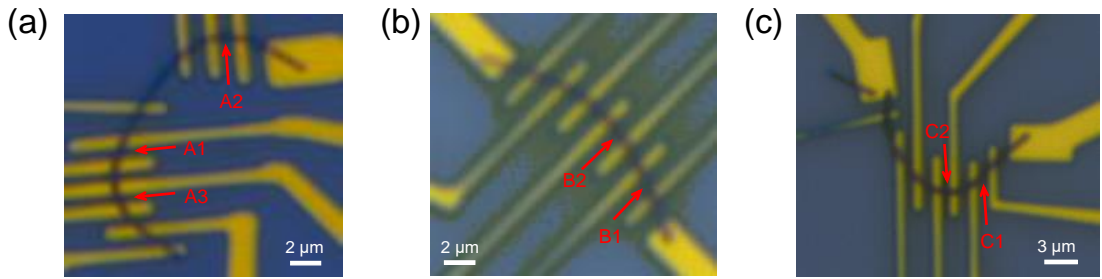

**FIG. 5. Optical images of device A, B and C, respectively.** The red arrows denote the bent regions with different curvature radius for each device.

Table 1: Geometric parameters of the nanoribbon devices.

| Number | device A | | | device B | | device C | |
|---|---|---|---|---|---|---|---|
| region | A1 | A2 | A3 | B1 | B2 | C1 | C2 |
| $r$ (μm) | 4.13 | 3.46 | 3.55 | 25.50 | 8.98 | 11.68 | 2.68 |
| $g$ (% $\mu m^{-1}$) | 24.2 | 28.9 | 28.2 | 3.9 | 11.1 | 8.6 | 37.3 |
| $\varepsilon$ (%) | 2.41 | 2.80 | 2.75 | 0.34 | 1.04 | 0.83 | 3.08 |

**APPENDIX B: TRANSPORT MEASUREMENTS ON ANOTHER BENT REGION OF DEVICE A.**

In device A, the bent region A3 has the similar curvature radius and maximal strain amplitude as those of region A2 described in Fig. 4. As expected, multiple-frequency oscillation pattern [Figs. 6(b) and 6(c)] is observed in the region A3. Figure 6(a) displays the evolution of resistivity as varying the temperatures. In striking contrast to case of bent region A1, the upturn below 10 K is missing in the region A3. Instead, the resistivity exhibits a monotonic decay as decreasing the temperature below 75 K. As seen from Fig. 2(e) in the main text, the large bending strain causes prominent band shift through the nanoribbon, where no global band gap exists and the sample Fermi level is always intersected with the energy bands. Thus, in the case, the insulating $\rho - T$ behavior is absent in the low temperature regime.

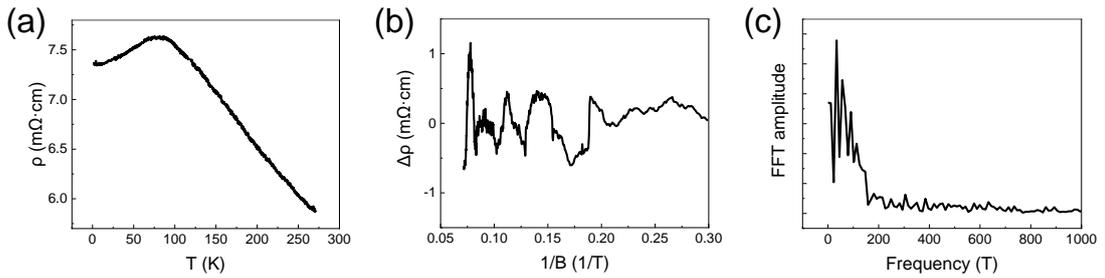

**FIG. 6. Transport measurements on the largely bent region A3 ($r = 3.55$ μm, $\varepsilon = 2.75\%$).** (a) Plot of resistivity $\rho$ versus temperature T. (b) The resistivity oscillation $\Delta\rho$ as a function of the inverse of the magnetic field $1/B$. (c) FFT spectrum of resistivity oscillation in (b).

**APPENDIX C: DIRAC SEMIMETAL PHASE IN THE RATHER WEAK**

# STRAIN REGIME AND DOUBLE-FREQUENCY OSCILLATION PATTERN

Figure 7 demonstrates the transport results measured on the device B. Upon a weak bending strain, the resulted lattice deformation is almost negligible and would not open a bulk gap, where the Dirac semimetal phase is still retained in the $Cd_3As_2$ nanoribbon. Figure 7(a) shows the SdH oscillations measured in the weakly bent region B1. The slope of linear plot in Landau level fan diagram gives the oscillation frequency of about 52 T [Fig. 7(b)]. The obtained intercept $-0.64$ is close to the theoretical value -5/8 expected in the Dirac semimetals [66]. Upon increasing the bending strain, the lattice deformation is gradually exacerbated and finally a band gap opens near the Dirac point. Resistivity oscillation pattern is also observed in the largely bent region B2, as shown in Fig. 7(c). Two peaks are found in the corresponding FFT spectrum [Fig. 7(d)]. The larger one (~56 T) is close to the value of oscillation frequency captured in region B1, which should arise from the cyclotron motion of bulk carriers. The smaller one (~34 T) is most likely to originate from the surface states of the emergent insulator phase.

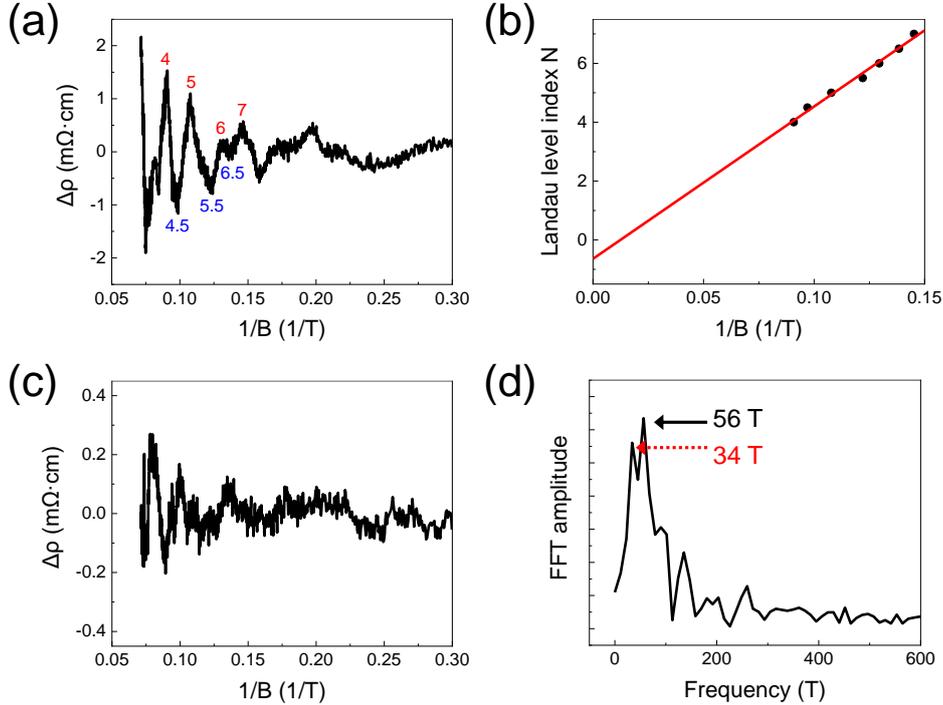

**FIG. 7. Magnetotransport measurements on the nanoribbon device B.** (a) SdH oscillations in the weakly bent region B1 ($r = 25.5$ μm, $\varepsilon = 0.34\%$). Landau level indices are labeled with oscillation peaks being integers and valleys being half-integers.

(b) Landau level fan diagram of SdH oscillations in (a). (c) The resistivity oscillations observed in the largely bent region B2 ($r = 8.98$ μm, $\varepsilon = 1.04\%$). (d) FFT spectrum of resistivity oscillation in (c).

**APPENDIX D: DISCUSSION ABOUT STRAIN EFFECTS ON THE TRANSPORT PROPERTIES OF BULK-DOMINATED NANORIBBONS.**

Generally, the electronic transport properties are closely associated with the energy band structure and the carrier density (the location of Fermi level) of the studied material. Figures 2(c-e) and related text describe the modulation of bending strain on the transport properties when the nanoribbon Fermi level is situated near the charge neutrality point. Here we make an analysis of the condition where the Fermi level is far away from the charge neutrality point [Fig. 8]. Under a magnetic field, the resistivity oscillations would transform from single-frequency to multiple-frequency pattern with increasing the strain strength. The single-frequency corresponds to the cyclotron motion of bulk carriers, different from the case of Fig. 2(d), where surface states contribute to the single-frequency pattern. Distinct origins may lead to different intercepts in the Landau level fan plot.

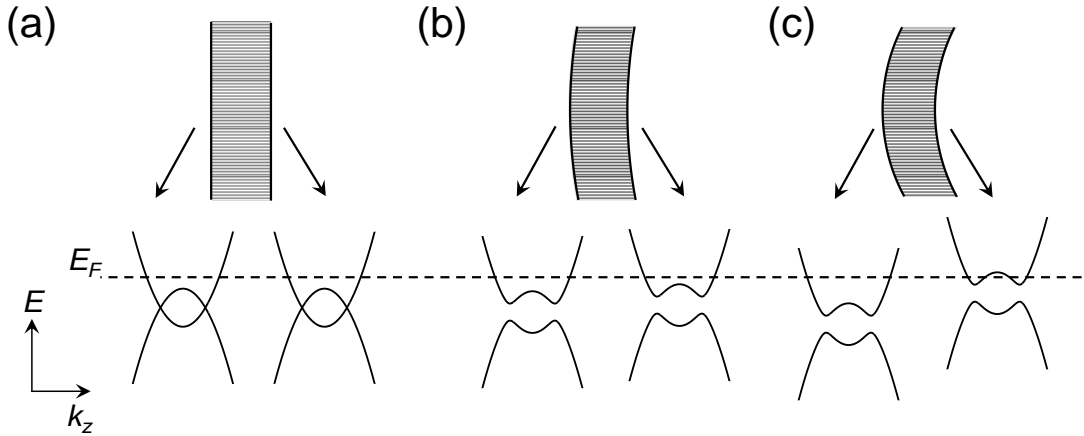

**FIG. 8. Schematic of bulk band structure evolution in the bent Cd$_3$As$_2$ nanoribbon as the bending strength increases.** Here the Fermi level of nanoribbon is assumed to be situated highly above the charge neutrality point, distinct from the case of Figs. 2(c-e).

# APPENDIX E: STRAIN-MEDIATED OSCILLATION PATTERNS IN THE HEAVILY ELECTRON DOPED NANORIBBON

Figure 9 shows the transport results measured on the bent nanoribbon device C. Seen from the transfer curve in Figs. 9(a) and 9(d), the nanoribbon is heavily electron doped and the Fermi level should situate deeply in the conduction bands. SdH oscillations are clearly observed in the weakly bent region [Fig. 9(b)], where oscillation peaks correspond to integer Landau level indices and oscillation valleys correspond to half-integers. The slope of Landau fan plot [Fig. 9(c)] shows the frequency of SdH oscillations, that is, $F = 72.06$ T. The large oscillation frequency further confirms that the Fermi level is far away from the charge neutrality point in this nanoribbon. The intercept ~0.9 is consistent with the theoretical value of topological semimetal near the Lifshitz transition point [66]. This can be easily understood since the strain is not enough to open a bulk gap, where the Dirac semimetal phase is still maintained in the region C1. Case is different when the applied bending strain is large. The local strain effect dominates and gives rise to spatially-dependent band shift [Fig. 8(c)]. Under a magnetic field, the cross-sectional area of "local Fermi surface" is non-uniform through the nanoribbon. Thus, multiple-frequency oscillation pattern is observed in the strongly-bent region C2, as shown in Figs. 9(e) and 9(f).

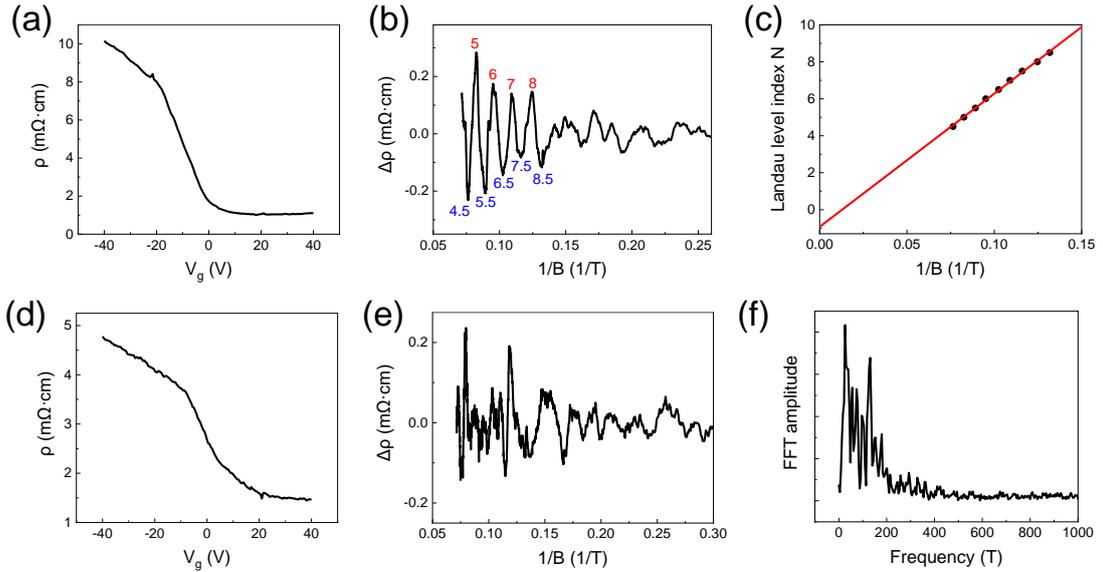

**FIG. 9. Transport measurements on the nanoribbon device C.** (a-c) correspond to the weakly bent region C1 ($r = 11.68$ μm, $\varepsilon = 0.83\%$), while (d-f) show the case of strongly bent region C2 ($r = 2.68$ μm, $\varepsilon = 3.08\%$). (a,d) Transfer curves of the bent region C1 and C2, respectively. (b) SdH oscillations in the region C1. Landau level indices are labeled with oscillation peaks being integers and valleys being half-integers. (c) Landau level fan diagram of SdH oscillations in (b). (e) Multiple-frequency resistivity oscillations observed in the region C2. (f) FFT spectrum of resistivity oscillation in (e).